\documentstyle[12pt,aps,prd,preprint,amssymb,tighten]{revtex}

\draft
\begin{document}
\author{Tom\'a\v s S\'ykora \thanks{%
e-mail address -- sykora@hp01.troja.mff.cuni.cz}}
\address{Institute of Particle and Nuclear Physics,
Faculty of Mathematics and Physics, Charles University,
V Hole\v sovi\v ck\'ach 2, 180 00 Prague, Czech Republic}
\title{Schwinger terms of the commutator of two interacting currents in the 1+1
dimensions}
%\date{1 September 1999}
\maketitle

\begin{abstract}
We calculate the equal-time commutator of two fermionic currents within the
framework of the 1+1 dimensional {\em fully} quantized theory, describing
the interaction of massive fermions with a massive vector boson. It is shown
that the interaction does not change the result obtained within the theory
of free fermions.
\end{abstract}

\section{Introduction}

\label{1} \renewcommand{\theequation}{1.\arabic{equation}} %
\setcounter{equation}{0}

When quantizing the fermionic currents there occur additional terms in the
equal-time (E.T.) commutators, so called Schwinger terms. They can be
determined by different methods, for instance perturbatively, see e.g.
\cite{Jackiw}-$\!\!\!$\cite{Banerjee}
%, \cite{Jo}, \cite{BertSykora},
or cohomologically
\cite{Faddeev,Mickelsson}, and they have an important impact on the
theory \cite{Jackiw}.

Schwinger terms are also closely related to the anomalies of QFT (for an
overview see \cite{Bertlmann}). Whereas anomalies do not get altered by
considering {\em quantized} gauge fields (due to the Adler-Bardeen theorem
\cite{AdlerBardeen})
this is less clear for the Schwinger terms (ST). Therefore it is our aim to
investigate ST in a {\em full} quantized theory. We will work in a 1+1
dimensional QFT describing the interaction of fermions with a massive vector
field. There all calculations can be performed explicitly and it is a
natural continuation of the work \cite{Sykora}, where the case of the free
fermions was discussed.

The fact that {\em all} fields are quantized distinguishes this work from
others where the similar calculations were done for the theories describing
fermions interacting with {\em external} fields (see e.g.
\cite{Ekstrand}-$\!\!$\cite{Hosono}
%\cite{Ekstrand,Adam,Hosono}
and the references given there).

There is one exception (known to the author) \cite{Hagen}, where boson field
is quantized too, but fermions are considered as {\em massless} and
the used procedure is quite different - from the begining the currents
are defined as the composite operators via the point-splitting
method.

The paper is organized as follows. In Section 2 we start with the definition
of the interacting current $J_{int}^\mu $\footnote{%
Label $int$ is to emphasize that the operator is {\em not} a composite
operator.} introduced by Bogoliubov \cite{Bogoljubov} in the framework of
the formalism of Epstein and Glaser \cite{Epstein} which is explained in
detail and extensively used in the book of Scharf \cite{Scharf}. Using this
definition we derive the explicit form of $J_{int}^\mu $ in the considered
two dimensional field theory model. The calculation of the commutator is
done in Section 3. In the Appendix we rigorously show that in the case of
our model, the formalism of Epstein and Glaser is equivalent to the ordinary
one using the ${\cal T}$-product (time-ordered product).

\section{Definition of the interacting current}

\label{2} \renewcommand{\theequation}{2.\arabic{equation}} %
\setcounter{equation}{0}

Following Bogoliubov \cite{Bogoljubov} and Scharf \cite{Scharf} we define
\begin{equation}
J^\mu (x)\equiv {\cal S}^{-1}(g)\left.\frac{\delta {\cal S}(g)}{i\delta g_\mu (x)}%
{}\right| _{g_\mu =0},  \label{DefIntCur}
\end{equation}
where the S-matrix ${\cal S}$ and its inverse ${\cal S}^{-1}$ are expressed
in perturbative form as
\begin{eqnarray}
{\cal S}(g) &&\equiv {\bf 1}+\sum_{n=1}^\infty \sum_{i=0}^n%
\frac{e^i}{i!\left( n-i\right) !}\times  \nonumber \\
&& \qquad \times \int T_n^{\mu _{1,}\ldots ,\mu _{n-i}}(x_1,\ldots ,x_n)g_{\mu
_1}(x_{i+1})\ldots g_{\mu _{n-i}}(x_n)\;d^2x_1\ldots d^2x_n
\label{DefS-Matrix} \\
&&  \nonumber \\
&&\equiv {\bf 1}+T, \\
&&  \nonumber \\
{\cal S}^{-1}(g) &&\equiv {\bf 1}+\sum_{n=1}^\infty
\sum_{i=0}^n\frac{e^i}{i!\left( n-i\right) !}\times  \nonumber \\
&& \qquad \times \int \tilde{T}_n^{\mu _{1,}\ldots ,\mu _{n-i}}(x_1,\ldots
,x_n)g_{\mu _1}(x_{i+1})\ldots g_{\mu _{n-i}}(x_n)\;d^2x_1\ldots d^2x_n,
\end{eqnarray}
where $e$ is the coupling constant of the interaction between fermions and
bosons and $g_{\mu _i}\left( x\right) $ is a c-number test functions from $%
{\cal S}\left({\mathbb{R}}^2\right) $ (Schwartz space). Properties of $%
T_n^{\mu _{1,}\ldots ,\mu _{n-i}}(x_1,\ldots ,x_n)$ are fixed by the
required properties of ${\cal S}$ (see \cite{Scharf}).

From the equation
\begin{equation}
{\cal S}(g)^{-1}=\left( {\bf 1}+T\right) ^{-1}={\bf 1+}\sum\limits_{r=1}^%
\infty \left( -T\right) ^r
\end{equation}
we get
\begin{equation}
\tilde{T}_n(X)=\sum\limits_{r=1}^n\left( -\right)
^r\sum\limits_{P_r}T_{n_1}\left( X_1\right) \ldots T_{n_r}\left( X_r\right) ,
\label{Definice T s vlnkou}
\end{equation}
where the second sum runs over all partitions $P_r$ of $X=\left\{
x_1,x_2,\ldots ,x_n\right\} $ into $r$ disjoint non-empty subsets.

For the meaningful definition of our field theory model we have to define
the first order terms of (\ref{DefS-Matrix})
\begin{equation}
{\cal S}^{(1)}(g)\equiv i\int \left\{ eT_1(x)+g_\mu (x)T_1^\mu (x)\right\}
d^2x,
\end{equation}
where
\begin{eqnarray}
T_1(x) &=&:\overline{\psi }(x){%
%TCIMACRO{\TeXButton{ \not{\! }\! A}{ \not{\! }\! A}}
%BeginExpansion
 \not{\! }\! A%
%EndExpansion
}\psi (x):\,,  \label{T1} \\
T_1^\mu (x) &=&:\overline{\psi }(x)\gamma ^\mu \psi (x):\,.  \label{T1mu}
\end{eqnarray}
The fields $\psi (x)$ and $\overline{\psi }(x)$ represent both fermion and
antifermion and the $A_\mu $ is a vector boson field. All fields appearing
in (\ref{T1}) and (\ref{T1mu}) are {\em free }since we work with a
perturbation expansion. The masses of the particles we denote
\begin{eqnarray*}
m_{\psi \left( \overline{\psi }\right) } &=&m, \\
m_A &=&\mu .
\end{eqnarray*}
Note that we use the model containing a massive vector field in order to
avoid problems with infrared singularities.

Using (\ref{DefIntCur}) and (\ref{DefS-Matrix}) we derive
\begin{equation}
J^\mu (x)=\,J_{free}^\mu (x)+\frac 1i\sum_{n=1}^\infty \frac 1{n!}\int
A_{n+1}^\mu (x_1,\ldots ,x_n;x)\;d^2x_1\ldots d^2x_n,
\label{tvar interagujiciho proudu}
\end{equation}
where
\begin{equation}
J_{free}^\mu (x)=:\overline{\psi }(x)\gamma ^\mu \psi (x):
\end{equation}
and $A_{n+1}^\mu $ is the so-called advanced $(n+1)$-point function
\begin{equation}
A_{n+1}^\mu (x_1,\ldots ,x_n;x)=\sum\limits_{P_2^0}\tilde{T}_m\left(
X/Y\right) T_{n-m}^\mu \left( Y,x\right) ,  \label{definice A}
\end{equation}
where $\sum_{P_2^0}$ means the summation over all partitions of the set $X$
including the empty subset $X/Y={%
%TCIMACRO{\TeXButton{ \not{\! } 0}{ \not{\! } 0}}
%BeginExpansion
 \not{\! } 0%
%EndExpansion
}$. The label {\em advanced} means that the support of $A_{n+1}^\mu $ is
\[
%TCIMACRO{\TeXButton{mbox}{\mbox{supp }}}
%BeginExpansion
\mbox{supp }%
%EndExpansion
A_{n+1}^\mu (x_1,x_2,\ldots ,x_n;x)\subseteq \Gamma _{n+1}^{-}\left(
x\right) ,
\]
where
\[
\Gamma _{n+1}^{-}\left( x\right) \equiv \left\{ \left. \left\{ x_i\right\}
_{i=1}^n\right| \left( x_i-x\right) ^2\geq 0,x_i^0\leq x^0\right\} ,
\]
i.e. the $A_{n+1}^\mu $ vanish if an arbitrary $x_i^0$ is greater then $x^0$.

For the reason which will become clear later we rewrite (\ref{definice A})
in the form
\begin{equation}
A_{n+1}^\mu (x_1,\ldots ,x_n;x)=\sum\limits_\Pi \,\theta \left(
x,x_{i_1},\ldots ,x_{i_n}\right) C_n\left( x,x_{i_1},x_{i_2},\ldots
,x_{i_n}\right) ,  \label{n bodova A}
\end{equation}
where
\begin{eqnarray}
C_n\left( x,x_{i_1},x_{i_2},\ldots ,x_{i_n}\right) &=&\left[ \ldots \left[
\left[ T_1^{\ \mu }\left( x\right) ,T_1\left( x_{i_1}\right) \right]
,T_1\left( x_{i_2}\right) \right] \ldots ,T_1\left( x_{i_n}\right) \right]
\nonumber \\
&&  \label{n komutator} \\
\theta \left( x,x_{i_1},\ldots ,x_{i_n}\right) &=&\theta \left(
x^0-x_{i_1}^0\right) \theta \left( x_{i_1}^0-x_{i_2}^0\right) \ldots \theta
\left( x_{i_{n-1}}^0-x_{i_n}^0\right)  \nonumber
\end{eqnarray}
and the summation runs over all permutations of the elements of $X$. (The
`advancing' of the support is then evident.)

Combining (\ref{tvar interagujiciho proudu}) and (\ref{n bodova A}) we get
\begin{equation}
J^\mu (x)=\frac 1i\sum\limits_{n=0}^\infty e^nJ_n^\mu (x),
\end{equation}
where
\begin{eqnarray}
J_0^\mu (x) &=&iJ_{free}^\mu (x)=T_1^{\ \mu }\left( x\right) \\
J_n^\mu (x) &=&\int \theta \left( x,x_1,x_2,\ldots ,x_n\right) C_n\left(
x,x_1,x_2,\ldots ,x_n\right) \;d^2x_1d^2x_2\ldots d^2x_n.
\label{n clen rozvoje int proudu}
\end{eqnarray}

Nevertheless the definition (\ref{DefIntCur}) has to be slightly modified if
we (naturally) require that the vacuum expectation value of the interacting
current $J^\mu (x)$ be equal to zero.

Our redefinition is then straightforward
\begin{eqnarray}
J_{int}^\mu (x) &\equiv &\,J^\mu (x)-\left\langle 0\right| J^\mu (x)\left|
0\right\rangle  \nonumber \\
&\equiv &\frac 1i\sum\limits_{n=0}^\infty e^n\left( J_n^\mu (x)-\left\langle
0\right| J_n^\mu (x)\left| 0\right\rangle \right)
\label{obecna formule pro proud}
\end{eqnarray}

\section{Commutator of interacting currents}

\label{3} \renewcommand{\theequation}{3.\arabic{equation}} %
\setcounter{equation}{0}

Now we are ready to calculate the commutator of two interacting currents.
Using (\ref{obecna formule pro proud}) we write
\begin{equation}
\left[ J_{int}^\mu (x),J_{int}^\nu (y)\right] _{E.T.}=\left[ J^\mu (x),J^\nu
(y)\right] _{E.T.}=-\sum\limits_{n=0}^\infty e^n\sum\limits_{i=0}^n\left[
J_i^\mu (x),J_{n-i}^\nu (y)\right] .
\end{equation}
and according to formulae (\ref{n clen rozvoje int proudu}), (\ref{n
komutator}) and due to the fact \cite{Schweber} that
\begin{eqnarray}
&& A_{n+1}^\mu (x_1,\ldots ,x_n;x)-A_{n+1}^\nu (x_1,\ldots
,x_n;y)\left. =\right.  \nonumber \\
&& \qquad = \sum\limits_{i_1\ldots i_n}\sum\limits_{k=0}^n\frac 1{%
k!\left( n-k\right) !}\left[ A_{n+1}^\mu (x_{i_1},\ldots
,x_{i_k};x),A_{n+1}^\nu (x_{i_{k+1}},\ldots ,x_{i_n};y)\right]  \nonumber \\
&&
\end{eqnarray}
we finally get

\begin{eqnarray}
&&\left[ J_{int}^\mu (x),J_{int}^\nu (y)\right] _{E.T.}\left. =\right.
\nonumber \\
&& \nonumber \\
&&\quad = -\sum\limits_{n=0}^\infty e^n\int \theta \left(
x,x_1,x_2,\ldots ,x_n\right) \times  \nonumber \\
&&\qquad \times \left[ \ldots \left[ \left[ T_1^{\ \mu }\left( x\right)
,T_1^{\ \nu }\left( y\right) \right] _{E.T.},T_1\left( x_1\right) \right]
\ldots ,T_1\left( x_n\right) \right] \;d^2x_1d^2x_2\ldots d^2x_n.  \nonumber
\\
&&
\end{eqnarray}
However, it turns out that only the first term in the sum contributes. To
see this, one has to realize that the {\em operator relation}
\begin{equation}
\left[ T_1^{\ \mu }\left( x\right) ,T_1^{\ \nu }\left( y\right) \right]
_{E.T.}\sim {\bf 1}  \label{ET operatorovy vztah}
\end{equation}
is valid. Indeed, using Wick theorem we express $\left[ T_1^\mu \left(
x\right) ,T_1^\nu \left( y\right) \right] $ in terms of the normally ordered
products
\begin{eqnarray}
\left[ T_1^\mu \left( x\right) ,T_1^\nu \left( y\right) \right]
&& = i\left. :\right. \overline{\psi }(y)\gamma ^\nu S\left(
y-x\right) \gamma ^\mu \psi (x)\left. :\right. -i\left. :\right. \overline{%
\psi }(x)\gamma ^\mu S\left( x-y\right) \gamma ^\nu \psi (x)\left. :\right. +
\nonumber \\
&&+\ \mbox{tr}\left\{ S^{(-)}\left( x-y\right) \gamma ^\nu S^{(+)}\left(
y-x\right) \gamma ^\mu \right. -\left. S^{(-)}\left( y-x\right) \gamma ^\mu
S^{(+)}\left( x-y\right) \gamma ^\nu \right\} .  \nonumber \\
&&
\end{eqnarray}
In the equal-time limit we have
\begin{equation}
S\left( x-y\right) \left. {}\right| _{E.T.}=\gamma ^0\delta \left(
x^1-y^1\right)
\end{equation}
and because in the 1+1 dimensions the identity
\begin{equation}
\gamma ^\mu \gamma ^0\gamma ^\nu =\gamma ^\nu \gamma ^0\gamma ^\mu
\end{equation}
holds, the relation (\ref{ET operatorovy vztah}) is proved\footnote{%
Here we are working with normal ordered operators, which are well-defined in
Fock space. Therefore we can indeed write
\begin{equation}
i:\overline{\psi }(y)\gamma ^\nu S\left( y-x\right) \gamma ^\mu \psi (x):-i:%
\overline{\psi }(x)\gamma ^\mu S\left( x-y\right) \gamma ^\nu \psi
(x):\left. {}\right| _{E.T.}=0,  \label{rozdil nekonecen}
\end{equation}
i.e. RHS of (\ref{rozdil nekonecen}) is {\em not} given as the difference of
two infinities
%TCIMACRO{\TeXButton{\cite{Kallen}}{\cite{Kallen}}}
%BeginExpansion
\cite{Kallen}.%
%EndExpansion
}.

Thus, we can conclude that no contribution from the interaction appears,
i.e.
\begin{equation}
\left[ J_{int}^\mu (x),J_{int}^\nu (y)\right] _{E.T.}=\left[ J_0^\mu
(x),J_0^\nu (y)\right] _{E.T.}  \label{hlavni vysledek}
\end{equation}
This is the main result of the article.

As the author has checked the same result can be obtained in the bosonization
scheme \cite{Abdalla}. Moreover, it can be shown that the work
\cite{Hagen} (where $m=0$) for the model considered here leads, in
fact, to the same answer. Therefore a connection between these
approaches and our scheme might exist.

\section{Conclusion}

\label{4} \renewcommand{\theequation}{4.\arabic{equation}} %
\setcounter{equation}{0}

We have calculated the commutator of interacting currents in the simple
two-dimensional model describing the interaction of fermions with a massive
vector field. We have shown that the interaction does not change the result
obtained within the theory of free fermions. A similar result we also expect
to occur in 3+1 dimensions, however, the same calculations cannot be carried
out in the same way in the 3+1 dimensions because the equality
\begin{equation}
\sum\limits_{P_2^0}\tilde{T}_m\left( X/Y\right) T_{n-m}^\mu \left(
Y,x\right) =\sum\limits_\Pi \,\theta \left( x,x_{\pi _1},\ldots ,x_{\pi
_n}\right) C_n\left( x,x_{\pi _1},\ldots ,x_{\pi _n}\right)
\end{equation}
is no longer valid.

Taking the limit $\mu \rightarrow 0$ for the vector field would lead to
problems because of the infrared singularities. Nevertheless these
singularities do not preclude the possibility to split the causal
distributions using multiplication by theta-functions. Therefore one may
guess that the relation (\ref{hlavni vysledek}) remains valid even in the
limit $\mu \rightarrow 0$.

The other types of couplings (e.g. chiral or axial) are under
consideration within the framework of our formalism.
The results might then be compared (in the limit $m=0$) with
those of the \cite{Hagen}.

In view of the intimate connection between Schwinger terms and anomaly, our
result naturally suggests another (open) question concerning its possible
relation to the Adler-Bardeen theorem. However, a {\em general} proof that
quantized gauge fields do not change the result of the external fields is
still missing.

\section*{Acknowledgments}

First of all my thanks belong to Dr. J. Novotn\'{y} and Dr. Ch. Adam
who independently showed me equivalence of my result with the result
of the work of Prof. C. R. Hagen in the limit $m=0$.

I thank Prof. C. R. Hagen who called my attention to his works.
I am grateful to Prof. R. Jackiw and Prof. R. Banerjee for useful
correspondence. Also I would like to express my thanks to
Prof. J. Ho\v {r}ej\v {s}\'{\i}, Prof. R. A. Bertlmann and Prof. H. Grosse
for fruitful discussions and comments.
It is pleasure to thank Prof. R. A. Bertlmann for the possibility to spend
some time in the inviting and creative atmosphere of the Erwin Schroedinger
Institute in Vienna.

This work was partly supported by Austria-Czech Republic Scientific
collaboration, project KONTAKT 1999-8 and by the research grant No. GA\v {C}%
R-202/98/0506.

\section*{Appendix: The transition step}

\renewcommand{\theequation}{A.\arabic{equation}} \setcounter{equation}{0}

To justify the transition from (\ref{definice A}) to (\ref{n bodova A}) we
introduce the distribution $D_{n+1}^\mu $
\begin{eqnarray}
D_{n+1}^\mu (x_1,\ldots ,x_n;x) &\equiv &R_{n+1}^\mu (x_1,\ldots
,x_n;x)-A_{n+1}^\mu (x_1,\ldots ,x_n;x),  \nonumber \\
&&  \label{definice D}
\end{eqnarray}
where
\begin{equation}
R_{n+1}^\mu (x_1,\ldots ,x_n;x)\equiv \sum\limits_{P_2^0}T_{n-m}^\mu \left(
Y,x\right) \tilde{T}_m\left( X/Y\right) .  \label{Definice R}
\end{equation}
It is possible to show that
\begin{equation}
\mbox{supp }R_{n+1}^\mu (x_1,x_2,\ldots ,x_n;x)\subseteq \Gamma
_{n+1}^{+}\left( x\right) ,
\end{equation}
where
\begin{equation}
\Gamma _{n+1}^{+}\left( x\right) \equiv \left\{ \left. \left\{ x_i\right\}
_{i=1}^n\right| \left( x_i-x\right) ^2\geq 0,x_i^0\geq x^0\right\}
\end{equation}
and therefore $D_{n+1}^\mu $ has a causal support, i.e.
\begin{equation}
\mbox{supp }D_{n+1}^\mu \left( X,x\right) \subseteq \Gamma _{n+1}^{+}\left(
x\right) \cup \Gamma _{n+1}^{-}\left( x\right) .
\end{equation}
It is clear that if $D_{n+1}^\mu $ is not singular then $A_{n+1}^\mu $ can
be expressed as
\begin{equation}
A_{n+1}^\mu (x_1,\ldots ,x_n;x)=-\prod_{i=1}^n\theta \left( x^0-x_i^0\right)
D_{n+1}^\mu (x_1,\ldots ,x_n;x).  \label{splitting}
\end{equation}
Further we introduce the `truncated' distributions $A_{n+1}^{\prime \ \mu }$
and $R_{n+1}^{\prime \ \mu }$%
\begin{eqnarray}
A_{n+1}^{\prime \mu } &\equiv &\sum\limits_{P_2}\tilde{T}_m\left( X/Y\right)
T_{n-m}^\mu \left( Y,x\right) ,  \label{A prime} \\
R_{n+1}^{\prime \ \mu } &\equiv &\sum\limits_{P_2}T_{n-m}^\mu \left(
Y,x\right) \tilde{T}_m\left( X/Y\right) ,  \label{R prime}
\end{eqnarray}
where $\sum_{P_2}$ means summation over all partitions of the set $X$ to
non-empty subsets. Using (\ref{A prime}) and (\ref{R prime}) the
distribution (\ref{definice D}) can be express as
\begin{equation}
D_{n+1}^\mu =R_{n+1}^{\prime \ \mu }-A_{n+1}^{\prime \ \mu }.
\label{nova definice D}
\end{equation}
There is one important difference between (\ref{definice D}) and (\ref{nova
definice D}). The latter gives us the possibility to express $D_{n+1}^\mu $
in terms of the $n$-point function $T^n$.

{\sl{
%\footnotesize
Example:}}{
%\footnotesize
{\
\begin{equation}
D_2^\mu (x_1;x)=R_2^{\prime \,\mu }(x_1;x)-A_2^{\prime \,\mu }(x_1;x)
\end{equation}
and if $D_2^\mu$ is not singular then we can write
\begin{equation}
A_2^\mu (x_1;x)=-\theta \left( x^0-x_1^0\right) D_2^\mu (x_1;x),  \label{D2}
\end{equation}
i.e. we split $D_2^\mu$. Then using (\ref{D2}) we get
\begin{eqnarray}
A_2^\mu (x_1;x) &=&-\theta \left( x^0-x_{\pi _1}^0\right) \left( R_2^{\prime
\ \mu }(x_1;x)-A_2^{\prime \ \mu }(x_1;x)\right) =  \nonumber \\
&=&\theta \left( x^0-x_{\pi _1}^0\right) \left[ T_1^\mu (x),T_1(x_1)\right] .
\end{eqnarray}
} }

Furthermore according to the definitions (\ref{definice A}), (\ref{A prime}), (%
\ref{Definice R}) and (\ref{R prime})
\begin{equation}
T_n^\mu \left( x_1,\ldots ,x_{n-1},x\right) =R_n^\mu -R_n^{\prime \ \mu
}=A_n^\mu -A_n^{\prime \ \mu }
\end{equation}
and using repeatedly (\ref{splitting}) we can finally express $D_{n+1}^\mu $
in the terms of the $1$-point function $T_1^{(\mu )}$. In that way we get $%
A_{n+1}^\mu $ in (\ref{n bodova A}) by combining (\ref{splitting}) and the
above procedure.

This routine is equivalent \cite{Schweber} to the application of following
equality
\[
T_{n+1}^\mu \left( x_1,\ldots ,x_{n-1},x\right) ={\cal T}\left( T_1^\mu
(x)T_1(x_1)\ldots T_1(x_n)\right) ,
\]
where
\begin{equation}
{\cal T}\left( T_1^\mu (x)T_1(x_1)\ldots T_1(x_n)\right) =\sum\limits_\Pi
\theta \left( x,x_{i_1},\ldots ,x_{i_n}\right) T_1^\mu (x)T_1(x_{i_1})\ldots
T_1(x_{i_n}).  \label{T product}
\end{equation}
{\sl{
%\footnotesize
Example:}} {
%\footnotesize
\begin{eqnarray}
T_2^\mu \left( x_1,x\right) &=&-\theta \left( x^0-x_1^0\right) D_2^\mu
-A_2^{\prime \,\mu }=  \nonumber \\
&=&-\theta \left( x^0-x_1^0\right) R_2^{\prime \,\mu }-\left( 1-\theta
\left( x^0-x_1^0\right) \right) A_2^{\prime \,\mu }=  \nonumber \\
&=&\theta \left( x^0-x_1^0\right) T_1^\mu (x)T_1(x_1)+\theta \left(
x_1^0-x^0\right) T_1^\mu (x)T_1(x_1)=  \nonumber \\
&=&T\left( T_1^\mu (x)T(x_1)\right).
\end{eqnarray}
}

However, as it was shown in \cite{Scharf} the splitting of an arbitrary
distribution with the causal support to retarded and advanced part via
multiplication by the combination of theta-functions, i.e. (\ref{splitting})
is not generally a well defined procedure.

The reason why we can do it here is that we work in two dimensions. We show
that the terms $T_{i_1}\ldots T_{i_k},i_j\in \{1,\ldots ,n\},$ $%
\sum_{j=1}^ki_j=n+1,$which are `sitting' in $D_{n+1}$, are correctly defined
and they have non singular behavior. The last enables their multiplication
by the combination of the theta-functions.

Every term $T_{i_1}\ldots T_{i_k}$ is expressible as a sum of terms of the
normally ordered operators (graphs) of the form
\begin{equation}
T_{i_1}\ldots T_{i_k}\thicksim \sum\limits_kT_{n+1}^{g_k}(x_1,\ldots ,x_n,x)
\end{equation}
where
\begin{equation}
T_n^g(x_1,\ldots ,x_n)=:\prod\limits_{i=1}^{n_f}\overline{\,\psi }%
(x_{k_j})t_g\left( x_{1,}\ldots ,x_n\right) \prod\limits_{i=1}^{n_f}\,\psi
(x_{n_j})::\prod\limits_{i=1}^{n_b}\,A\left( x_{m_j}\right) :  \label{graf}
\end{equation}
and $n_f$ is the number of external fermions (or antifermions), $n_b$ the
number of the external massive bosons and $t_g\left( x_{1,}\ldots
,x_n\right) $ is c-number distribution.

In the dimension $d$ the graph $g$ (\ref{graf}) has the singular order
\begin{equation}
\omega \left( g\right) =n\left( \frac d2-2\right) +d-n_b\left( \frac d2%
-1\right) -n_f\left( d-1\right)
\end{equation}
and for $d=2$ we get
\begin{equation}
\omega \left( g\right) =2-n-n_f.
\end{equation}
We see that the problematic (singular) case $\omega \left( g\right) \geq 0$
can appear only for $n=2$. All higher-order graphs do not contain any
singularity. Moreover the c-number distribution in the $2$-point causal
function has really $\omega \left( g\right) =-2$. This all means that all
graphs (including their subgraphs) are not singular, the terms $T_{i_1}\ldots
T_{i_k}$ are well-defined and can be multiplied by the combination of
theta-functions.

Therefore the formula (\ref{n bodova A}) is consistent
with (\ref{definice A}).


\begin{references}
\bibitem{Jackiw}  R. Jackiw, Field theoretic investigations in current
algebra, Topological investigations of quantized gauge theories, in: Current
algebra and anomalies, S.B. Treiman, R. Jackiw, B. Zumino and E. Witten (eds.), p. 81 and
p. 211, World Scientific, Singapore 1985.

\bibitem{Jo}  G. Jo, Phys. Rev. D35 (1987) 3179, Nucl. Phys. B259 (1985) 616.

\bibitem{BertSykora}  R. Bertlmann and T. S\'{y}kora, Phys. Rev. D56 (1997)
2236.

\bibitem{Banerjee}  R. Banerjee and S. Ghosh, Z. Phys C41 (1988) 121; Phys.
Lett. 220B (1989) 581; Mod. Phys. Lett A5 (1989) 855.

\bibitem{Faddeev}  L. Faddeev, Phys. Lett. 145B (1984) 81; L. Faddeev and S.
Shatashvili, Theor. Math. Phys. 60 (1984) 770.

\bibitem{Mickelsson}  J. Mickelsson, Commun. Math. Phys. 97 (1985) 361.

\bibitem{Bertlmann}  R. Bertlmann, Anomalies in quantum field theory,
International Series of Monographs on Physics 91, Clarendon -- Oxford
University Press, 1996.

\bibitem{AdlerBardeen}  S. L. Adler, W. A. Bardeen, Phys. Rev.
182, 1517.

\bibitem{Sykora}  T. S\'{y}kora, Schwinger terms in 1+1
dimensions, Czech. J. Phys, c. 6/99, vol. 49, p. 915-932.

\bibitem{Ekstrand}  Ch. Ekstrand, Schwinger Terms from External
Field Problems, Doctoral Dissertation, Royal Institute of Technology,
Department of Physics, Stockholm 1999.

\bibitem{Adam}  C. Adam, Consistent and Covariant Commutator
Anomalies in the Chiral Schwinger Model, hep-th/9710042 or MIT-CTP-2662,
September 1997.

\bibitem{Hosono} S. Hosono, Nucl. Phys. B300 (1988) 238-252

\bibitem{Hagen}  C. R. Hagen, Phys. Rev. D55 (1997) 1021;
Nuovo Cimento B 51 (1967) 169; Nuovo Cimento A 51 (1967) 1033;
Ann. Phys. 81 (1973) 67.

\bibitem{Bogoljubov}  N.N. Bogoliubov, D.V. Shirkov,
Introduction to the Theory of Quantized Fields, New York 1959.

\bibitem{Epstein}  H. Epstein and V. Glaser, Ann. Inst. Poincar%
%TCIMACRO{\TeXButton{TeX field}{\'e \ } }
%BeginExpansion
\'e \ %
%EndExpansion
A 19 (1973) 211.

\bibitem{Scharf}  G. Scharf, Finite Quantum Electrodynamics -
The Causal Approach, second edition, ISBN 3-540-60142-2, Springer-Verlag
Berlin Heidelberg New York 1989, 1995.

\bibitem{Schweber}  S. S. Schweber, An Introduction to
Relativistic Quantum Theory, Row, Peterson and Co, Evanston, Ill. Elmsford,
New York, 1961 (russion translation: GILL, Moscow 1963, p. 709-713).

\bibitem{Kallen}  G. K\"{a}llen, Gradient terms in commutators
of currents and fields, Lectures given at winter schools in Karpacz and
Schladming, February and March 1968.

\bibitem{Abdalla}  E. Abdalla, M. C. B. Abdalla and K. D.
Rothe, Non-Perturbative Methods in 2 Dimensional Quantum Field Theory,
Singapore World Scientific 1991.
\end{references}
\end{document}